\documentclass[final]{elsarticle}

\usepackage{hyperref}

\journal{be approved.}

\usepackage{amsthm,bm}

\usepackage{amssymb}
\usepackage{amsmath}
\usepackage{natbib}
\usepackage{amsmath,amssymb}
\usepackage{stackengine}
\usepackage{caption} 
\usepackage{graphicx}
\usepackage{booktabs}
\usepackage{enumerate}
\usepackage{algorithm}
\usepackage{algpseudocode}
\usepackage{lscape}
\usepackage{subcaption,dcolumn}
\usepackage{multirow}

\graphicspath{ {./Diagram/} }

\DeclareMathOperator{\PR}{\mathbb{P}}
\DeclareMathOperator{\E}{\mathbb{E}}

\makeatletter
\newcommand{\StatexIndent}[1][3]{%
	\setlength\@tempdima{\algorithmicindent}%
	\Statex\hskip\dimexpr#1\@tempdima\relax}
\algdef{S}[WHILE]{WhileNoDo}[1]{\algorithmicwhile\ #1}%
\makeatother

\DeclareUnicodeCharacter{2212}{-}

\let\-=\mathbf
\begin{document}

\begin{frontmatter}

\title{Multicanonical Sequential Monte Carlo Sampler for Uncertainty Quantification}

\author{Robert Millar\corref{cor1}}
\ead{r.millar.1@bham.ac.uk}

\author{Jinglai Li}
\ead{j.li.10@bham.ac.uk}

\author{Hui Li}
\ead{h.li.4@bham.ac.uk}

\address{School of Mathematics, University of Birmingham, Birmingham B15 2TT, UK}

\begin{abstract}
	In many real-world engineering systems, the performance or reliability of the system is characterised by a scalar parameter. The distribution of this performance parameter is important in many uncertainty quantification problems, ranging from risk management to utility optimisation. In practice, this distribution usually cannot be derived analytically and has to be obtained numerically by simulations. To this end, standard Monte Carlo simulations are often used, however, they cannot efficiently reconstruct the tail of the distribution which is essential in many applications. One possible remedy is to use the Multicanonical Monte Carlo method, an adaptive importance sampling scheme. In this method, one draws samples from an importance sampling distribution in a nonstandard form in each iteration, which is usually done via Markov chain Monte Carlo (MCMC). MCMC is inherently serial and therefore struggles with parallelism. In this paper, we present a new approach, which uses the Sequential Monte Carlo sampler to draw from the importance sampling distribution, which is particularly suited for parallel implementation. With both mathematical and practical examples, we demonstrate the competitive performance of the proposed method.
\end{abstract}


\cortext[cor1]{Corresponding author}

\end{frontmatter}

\section{Introduction}
Real-world engineering systems are unavoidably subject to uncertainty, rising from various sources: material properties, geometric parameters, external perturbations and so on.
In practice, it is vital to characterise and quantify the impact of the uncertainties on the system performance or reliability, which constitutes a central task in the field of uncertainty quantification (UQ).
Mathematical models and simulations are important tools to assess how engineering systems are impacted by the uncertainty. Within these, the system  performance or reliability is often characterised by a scalar parameter \emph{y}, which we will now refer to as the \emph{performance variable}. This performance variable can be expressed by a performance function $y = g(\mathbf{x})$, where $\mathbf{x}$ is a multi-dimensional 
random variable representing all the uncertainty factors affecting the system;
the performance function  is usually not of analytical form, 
and needs to be evaluated by simulating the underlying mathematical model.
A typical example is in structural engineering, where the performance variable $y$ is the deformation of some key components. 
The distribution of this performance variable is important in many UQ problems, ranging from risk management to utility optimisation. 
A challenge here is that these UQ problems may demand various statistical information of the performance $y$:
for example, in robust optimisation, the interests are predominantly in the mean and variance~\cite{du2004sequential},
in risk management, one is interested in the tail probability as well as some extreme quantiles~\cite{rockafellar2000optimization},
and in utility optimisation, the complete distribution of the performance parameter is required~\cite{hazelrigg1998framework}.  
To this end, methods that can efficiently reconstruct the probability distribution of the performance variable directly are strongly desirable. 
In principle the distribution of $y$ can be estimated by standard Monte Carlo (MC) simulations, however MC can be prohibitively expensive for systems with complex mathematical models.
In our previous works~\cite{wu2016surrogate,chen2017subset}, we proposed using the Multicanonical Monte Carlo (MMC) method for computing the distribution of $y$. 
The MMC method is a special adaptive importance sampling (IS) scheme, which was initially developed by Berg and Neuhaus~\cite{berg1991multicanonical,berg1992multicanonical} to explore the energy landscape of a given physical system.

In the MMC method, one splits the state space of the performance parameter of interest into a set of small bins and then iteratively constructs a so-called flat-histogram distribution that can assign equal probability to each of the bins. This allows for the construction of the entire distribution function of the performance parameter, significantly more efficiently than using standard MC.
There are other advanced MC techniques developed in reliability engineering, such as the cross-entropy method~\cite{li2011efficient}, subset simulation~\cite{AuB99}, sequential Monte Carlo~\cite{cerou2012sequential}, etc.
These methods are designed to provide a variance-reduced estimator for a specific quantity associated with the distribution of $y$, such as the probability of a given random event,  rather than reconstructing the distribution itself.

A key characteristic of MMC is that, within each iteration, samples are drawn from an IS distribution in a nonstandard form, which is usually done via Markov chain Monte Carlo (MCMC). MCMC is inherently serial \cite{hafych2022parallelizing}, in that it relies on the convergence of a single Markov chain to its stationary distribution,
and therefore often struggles with parallelism. As a result 
the MMC method implemented with MCMC (referred to as
MMC-MCMC hereafter) cannot take advantage of high-powered parallel computing. There are further limitations to MCMC - detailed in Section \ref{MCMCvSMCS} - which reduce the overall efficiency of the MMC-MCMC method. We propose using an alternative sampling method, namely the Sequential Monte Carlo sampler (SMCS), to draw samples from the IS distributions. 
The SMCS method, first developed in \cite{del2006sequential}, can fulfil the same role as MCMC in that, by conducting sequential IS for a sequence of intermediate distributions, it can generate (weighted) samples from an arbitrary target distribution.
The reason that we choose to implement MMC with the SMCS method is two-fold: first, since SMCS is essentially an IS scheme, it is easily parallelisable; second, SMCS can take advantage of a sequence of intermediate distributions, allowing it to be effectively integrated into the MMC scheme. Both points will be elaborated on later.

The rest of the paper is organized as follows. In Section \ref{Section:MMC}, we present the Multicanonical Monte Carlo method and in Section \ref{SMCSDetail}, the Sequential Monte Carlo sampler. We bring these techniques together in Section \ref{Section:MMCSMCS} to present the proposed \emph{Multicanonical Sequential Monte Carlo Sampler} and then apply this to various numerical examples in Section \ref{Section:NumEx}. Finally, Section \ref{Section:Conclusion} provides concluding remarks.

\section{Multicanonical Monte Carlo method}\label{Section:MMC}
\subsection{Problem setup and the Monte Carlo estimation}
We start with a generic setup of the problems considered here.
 Let $\mathbf{x}$ be a $d$-dimensional random vector following distribution $p_\mathbf{x}(\cdot)$, and let $y$ be a scalar variable characterised by 
 a function $y = g(\mathbf{x})$.
We want to determine the probability density function (PDF) of $y$, given by $\pi(y)$, where we assume that both $\mathbf{x}$ and $y$ are continuous random variables.

\begin{figure}
		\centering
		\includegraphics[width=100mm]{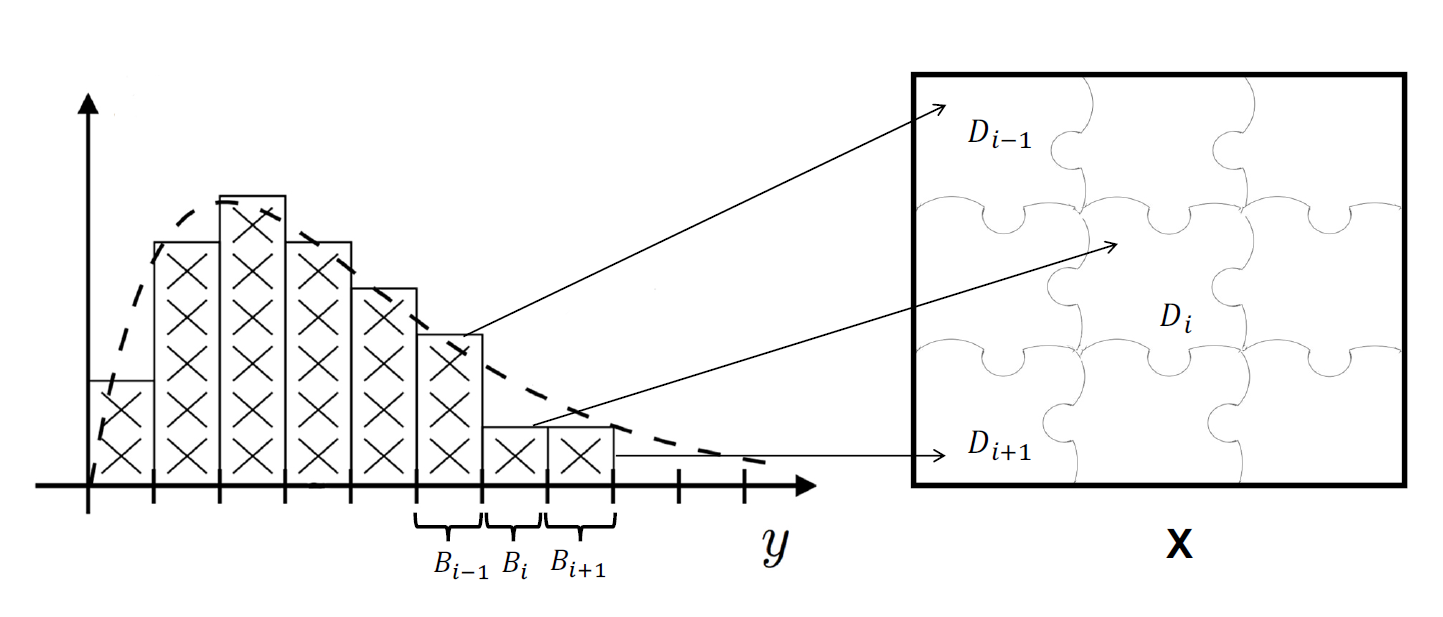}
		\caption{Schematic illustration of the connection between $B_i$ and $D_i$. This figure is reprinted from \cite{wu2016surrogate}.}
		\label{fig:MultiBin}		
	\end{figure}

We now discuss how to estimate the PDF using the standard MC simulation.
For the sake of convenience, we assume that $\pi(y)$ has a bounded support $R_y=[a,b]$, and if the support of $\pi(y)$ is not bounded, we choose the interval $[a,b]$ that is sufficiently large so that $\mathbb{P}(y\in[a,b])\approx1$.
We first decompose $R_y$ into $M$ bins of equal width $\Delta$ centred at the discrete values $\{b_1,...,b_M\}$, and define the $i$-th bin as the interval $B_i = [b_i - \Delta/2, b_i + \Delta/2]$.
This binning implicitly defines a partition of the input space $X$ into $M$ domains $\{D_i\}_{i=1}^{M}$, where
    \begin{equation}
		D_i = \{\mathbf{x} \in X : g(\mathbf{x}) \in B_i\}
	\end{equation}
is the domain in X that maps into the $i$-th bin $B_i$ (see Fig.~\ref{fig:MultiBin}).

While $B_i$ are simple intervals, the domains $D_i$ are multidimensional regions with possibly tortuous topologies. Therefore, an indicator function is used to classify whether a given $\mathbf{x}$-value is in the bin $D_i$ or not. Formally, the indicator function is defined as,
	\begin{equation}
		I_{D_i}(\mathbf{x}) = \begin{cases}
			1, & \text{if $\mathbf{x} \in D_i$};\\
			0, & \text{otherwise}
		\end{cases}
	\end{equation}
or equivalently $\{y = g(\mathbf{x}) \in B_i\}$.
By using this indicator function, the probability  that $y$ is in the $i$-th bin, i.e. $P_i = \PR\{y \in B_i\}$, can be written as an integral in the input space:
	\begin{equation}
		P_i = \int_{D_i}p(\mathbf{x})d\textbf{x} = \int I_{D_i}(\mathbf{x})p(\mathbf{x})d\textbf{x} = \E[I_{D_i}(\mathbf{x})].
		\label{Pi_estimator}
	\end{equation}
We can estimate $P_i$ via a standard MC simulation. 
Namely, we draw $N$ i.i.d. samples $\{\mathbf{x}_1,...,\mathbf{x}_N\}$ from the distribution $p(\mathbf{x})$,
and calculate the MC estimator of $P_i$ as 
\begin{equation}
		\hat{P}_i^{MC} = \frac{1}{N} \sum_{j=1}^{N} I_{D_i}(\mathbf{x}^j) = \frac{N_i}{N},\quad \mathrm{for}\,\, i=1,\,...,M,
		\label{MC_estimator}
	\end{equation}
where $N_i$ is the number of samples that fall in bin $B_i$.

Once we have obtained  $\{P_i\}_{i=1}^M$,  the PDF of $y$ at the point $y_i\in B_i$ - for a sufficiently small $\Delta$ - can be calculated as $\pi(y_i) \approx P_i / \Delta$.

\subsection{Flat Histogram Importance Sampling}
The MC approach can be improved through the use of Importance Sampling. Here IS is used to artificially increase the number of samples falling in the tail bins of the histogram. Given an IS distribution $q(\mathbf{x})$, Eq. \ref{Pi_estimator} can be re-written as
	\begin{equation}
		P_i = \int I_{D_i}(\mathbf{x})[\frac{p(\mathbf{x})}{q(\mathbf{x})}]q(\mathbf{x})d\textbf{x} = \E_q[I_{D_i}(\mathbf{x})w(\mathbf{x})]
	\end{equation}
where $w(\mathbf{x}) = p(\mathbf{x})/q(\mathbf{x})$ is the IS weight and $\E_q$ indicates expectation with respect to the IS distribution $q(\mathbf{x})$.
 The IS estimator for $P_i$ can then be written as follows:
	\begin{equation}
		\hat{P}_i^{IS} =  \left[\frac{1}{N} \sum_{j=1}^{N} I_{D_i}(\mathbf{x}^j) w(\mathbf{x}^j)\right]
		\label{IS_estimator}
	\end{equation}
for each bin $i=1,...,M$.

As is well known, key to the successful implementation of IS is identifying a good IS distribution $q(\mathbf{x})$, which is particularly challenging for the present problem, as we are interested in multiple estimates (i.e. $P_1,\,...,\,P_M$) rather than a single one, as in conventional IS problems.

The solution provided by MMC is to use the so-called \emph{uniform weight flat-histogram (UW-FH)} IS distribution. 
The UW-FH IS distribution  is designed to achieve the following two goals.
First, it should allocate the same probability to each bin, i.e. assuming $x\sim q(x)$, 
$$P_i^{\ast}:=\mathbb{P}(y=g(\mathbf{x})\in B_i) = 1/M,$$ for all $i$. Intuitively, this property allows all bins to be equally visited by the samples generated from the IS distribution. 
Second,  it should assign a constant weight to all samples falling in the same bin, 
that is, $w(\mathbf{x})  = \Theta_{i}$ for all $\mathbf{x} \in D_i$,
where $\Theta_i$ is a positive constant. 
Loosely speaking, the second property ensures that all samples falling in the same bin are equally good.

The UW-FH distribution can be expressed in the form of:
	\begin{equation}\label{e:uwfh}
		q(\mathbf{x}) \propto \begin{cases}
			\frac{p(\mathbf{x})}{c_\Theta\Theta(\mathbf{x})}, & \mathbf{x} \in D,\\
			0, & \mathbf{x} \notin D,
		\end{cases}
	\end{equation}
where $\Theta(\mathbf{x}) = \Theta_{i}$ for $ \mathbf{x} \in D_i,\; i = 1,...,M$,
and $c_\Theta$ is a normalizing constant. It is easy to see that, 
	\begin{equation}
		P_i^{\ast} = \int_{D_i} q(\mathbf{x}) d\mathbf{x} =\frac{\int_{D_i} p(\mathbf{x})d\mathbf{x}}{c_\Theta\Theta_i} = \frac{P_i}{c_\Theta\Theta_i}.
		\label{biasing_PMF}
	\end{equation}
Recall that $P_i^*=1/M$ for all $i$, so it follows $\Theta_i \propto P_i$, i.e. $\Theta_i$ is proportional to the sought probability $P_i$,
and  $c_{\Theta} = \sum_{i=1}^{M} \frac{P_i}{\Theta_i}$.
		
\subsection{Multicanonical Monte Carlo}
The UW-FH  distribution, given by Eq.~\eqref{e:uwfh}, cannot be used directly as $\Theta_{i}$ depends on the sought-after unknown $P_i$. The MMC method iteratively addresses this, starting from the original input PDF $p(\mathbf{x})$.
	
Simply put, starting with $q_{0}(\mathbf{x})$ and $\Theta_{0,i} = p$ for all $i = 1,...,M$, where $p = \sum_{i=1}^{M}P_i$, the MMC method iteratively constructs a sequence of distributions (for $t\geq1$)
	\begin{equation}\label{e:qt}
		q_t(\mathbf{x}) \propto \begin{cases}
			\frac{p(\mathbf{x})}{c_t\Theta_{t}(\mathbf{x})}, & \mathbf{x} \in D;\\
			0, & \mathbf{x} \notin D.
		\end{cases}
	\end{equation}
where $\Theta_t(\mathbf{x}) = \Theta_{t,i}$ for $\mathbf{x} \in D_i$ and $c_t$ is the normalizing constant for $q_t$. 
 Ideally we want to construct $q_t$ in a way that it converges to the actual UW-FH distribution
as $t$ increases. 
The key here is to estimate the values of $\{ \Theta_{t,i}\}_{i=1}^M$.
It is easy to see that when $q_t$ is used as the IS distribution, we have $P_i = c_t P_{i}^{\ast}\Theta_{t,i}$.

That is, in the $t$-th iteration, one draws $N$ samples $\{\mathbf{x}^j\}^N_{j = 1}$ from the current IS distribution $q_t(\mathbf{x})$, then updates  $\{\Theta_{t+1,i}\}_{i=1}^M$  using the following formulas,
	\begin{subequations}
		\label{e:params}
		\begin{gather}
			\hat{H}_{t,i} = \frac{N_{t,i}^{\ast}}{N}\label{e:Hti}\\
			P_{t,i} = \hat{H}_{t,i} \; \Theta_{t,i}\label{e:Pti}\\
			\Theta_{t+1,i} = P_{t,i}
		\end{gather}
	\end{subequations}
where $N_{t,i}^{\ast}$ is the number of samples falling into region $D_i$ in the $t$-th iteration. 
Note that in Eq.~\eqref{e:Pti} we neglect the normalizing constant $c_t$ as it is not needed in the algorithm, which will become clear later. 
The process is then repeated, until the resulting histogram is sufficiently ``flat'' (see e.g. \cite{iba2014multicanonical}).

\subsection{The limitation of MCMC}
\label{MCMCvSMCS}
To implement the MMC method, one must be able to generate samples from the IS distribution $q_t(\cdot)$ at each iteration. Typically, this is done using Markov Chain Monte Carlo (MCMC).
Simply speaking, MCMC constructs a Markov Chain that converges to the target distribution. It is convenient to use as it only requires the ability to evaluate the target PDF up to a normalizing constant (and therefore the knowledge of $c_t$
in Eq.~\eqref{e:qt} is not needed). 
The core of MCMC is to construct a single Markov chain converging to its stationary distribution, which often takes a very large number of iterations (known as the burn-in period) to be achieved. The process cannot be easily accelerated by parallel processing. 
We note here that there are some MCMC variants, e.g. \cite{vanderwerken2013parallel}, that attempt to exploit parallel implementation; however, to the best of our knowledge, none of these methods can take full advantage of modern parallel computing power.
For example, the multi-chain MCMC algorithms can be implemented in parallel, but each single chain still requires a long burn-in period before it converges to the target distribution. 
As a result, MMC-MCMC cannot fully exploit the potential provided by high-performance parallel computing available nowadays.
In this work, we want to provide an alternative implementation of MMC, which is based on the sequential Monte Carlo sampler.

\section{Sequential Monte Carlo sampler} \label{SMCSDetail}
First proposed in \cite{del2006sequential}, SMCS is an IS method for drawing samples from a sequence of distributions $\{q_t(\cdot)\}_{t=1}^{T}$. It is a generalisation of the particle filter \cite{arulampalam2002tutorial}, where weighted samples are generated 
in a sequential manner. Several extensions to this method have been proposed, e.g. \cite{beskos2017multilevel,heng2020controlled,green2022increasing,south2019sequential}, with the latest advances being summarised in two recent reviews \cite{chopin2020introduction, dai2020invitation}.
	
Suppose we have samples following distribution $q_{t-1}(\cdot)$ but want them to follow $q_t(\cdot)$ instead, we can use SMCS. First, a forward Kernel is applied to each of the current samples - sometimes with an acceptance criterion - and then a weight is calculated for each new sample. Finally, if the effective sample size across all the samples is below a certain threshold (usually less than half the total number of samples) the proposed samples are resampled. These new weighted samples follow the distribution $q_t(\cdot)$.

We present the SMCS method in a recursive formulation, largely following the presentation of \cite{del2006sequential} and \cite{wu2020ensemble}.
Suppose that at time $t-1$, we have an IS distribution $\gamma_{t-1}(\mathbf{x}_{t-1})$,
from which we have or can generate an ensemble of $N$ samples $\{\mathbf{x}_{t-1}\}_{j=1}^N$. 
To implement SMCS, we first choose two conditional distributions $K_t(\cdot|x_{t-1})$ and $L_{t-1}(\cdot|x_t)$, referred to 
as the forward and backward kernels respectively. 
Using $L_{t-1}(\cdot|x_{t})$, we are able to construct a joint distribution of $\mathbf{x}_{t−1}$ and $\mathbf{x}_t$ in the form of
	\begin{equation}
		r_t(\mathbf{x}_{t−1}, \mathbf{x}_t) = q_{t}(\mathbf{x}_t)L_{t−1}(\mathbf{x}_{t−1}|\mathbf{x}_{t})
	\end{equation}
such that the marginal distribution of $r_t(\mathbf{x}_{t−1}, \mathbf{x}_t)$ over $\mathbf{x}_{t−1}$ is $q_t(\mathbf{x}_t)$. Now, using  $\gamma_{t-1}(\mathbf{x}_{t-1})$ and the forward Kernel $K_t(\mathbf{x}_t|\mathbf{x}_{t-1})$, we can construct an IS distribution for $r_t(\mathbf{x}_{t−1}, \mathbf{x}_t)$ in the form of
	\begin{equation} \label{SMCS_3.2}
		\gamma(\mathbf{x}_{t−1}, \mathbf{x}_t) = \gamma_{t−1}(\mathbf{x}_{t−1})K_t(\mathbf{x}_t|\mathbf{x}_{t−1}).
	\end{equation}
One can draw samples from 
this joint IS distribution $\gamma(\mathbf{x}_{t−1}, \mathbf{x}_t)$ using $\{\mathbf{x}_{t-1}\}_{j=1}^N$ and the forward kernel $K_t$,
and let $\{(\mathbf{x}^{j}_{t-1},\mathbf{x}^j_t)\}^{N}_{j=1}$ be an ensemble drawn from $\gamma(\mathbf{x}_{t−1}, \mathbf{x}_t)$.
The corresponding weights are computed as
	\begin{subequations} \label{SMCS_weight_std}
		\begin{equation}
			\begin{aligned} \label{SMCS_weight_full}
				w_t(\mathbf{x}_{t-1:t}) &= \frac{r_t(\mathbf{x}_{t−1}, \mathbf{x}_t)}{\gamma(\mathbf{x}_{t−1}, \mathbf{x}_t)} = \frac{q_{t}(\mathbf{x}_t)\;L_{t−1}(\mathbf{x}_{t−1}|\mathbf{x}_{t})}{\gamma_{t−1}(\mathbf{x}_{t−1})\;K_t(\mathbf{x}_t|\mathbf{x}_{t−1})}\\
				&= w_{t-1}(\mathbf{x}_{t-1})\alpha(\mathbf{x}_{t-1},\mathbf{x}_t)
			\end{aligned}
		\end{equation}
		where
		\begin{equation}
			\begin{aligned} \label{SMCS_weight_split}
				w_{t-1}(\mathbf{x}_{t-1}) &= \frac{q_{t-1}(\mathbf{x}_{t-1})}{\gamma_{t−1}(\mathbf{x}_{t−1})},\\
				\alpha_t(\mathbf{x}_{t-1},\mathbf{x}_t) &= \frac{q_{t}(\mathbf{x}_t)\;L_{t−1}(\mathbf{x}_{t−1}|\mathbf{x}_{t})}{q_{t−1}(\mathbf{x}_{t−1})\;K_t(\mathbf{x}_t|\mathbf{x}_{t−1})}
			\end{aligned}.
		\end{equation}
	\end{subequations}
As such the weighted ensemble $\{\mathbf{x}^{j}_{t-1:t},w^{j}_{t}\}^{N}_{j=1}$ follows the joint distribution $r_t(\mathbf{x}_{t−1:t})$, 
and as such, $\{\mathbf{x}^{j}_{t},w^{j}_{t}\}^{N}_{j=1}$ follows the marginal distribution $q_t$. By repeating this procedure we can obtain weighted samples from the sequence of distributions $\{q_t\}_{t=1}^T$.

For the SMCS method, the choice of forward and backward kernels are essential. 
While noting that there are a number of existing methods for determining the forward kernel,
we adopt the MCMC kernel proposed in \cite{del2006sequential}, which is closely related to the Metropolis step 
in MCMC as the name suggests. 
Specifically, the forward kernel (more precisely the  process for generating samples from the forward kernel) is constructed as follows. 
We chose a proposal distribution $k(\-x_t|\-x_{t-1})$, 
and with a sample from the previous iteration $\-x^j_{t-1}$, we draw a sample $\-x^*_t$  from $k(\-x_t|\-x^j_{t-1})$, and then accept (or reject) $\-x^*_t$ according to the following acceptance probability:
	\begin{equation}
		a_t(\mathbf{x}^{\ast}_{t}|\mathbf{x}^j_{t-1}) = \text{min}\left\{\frac{q_t(\mathbf{x}^{\ast}_{t})}{q_t(\mathbf{x}^j_{t-1})}
		\frac{k(\-x^j_{t-1}|\-x_t^*)}{k(\-x^*_{t}|\-x^j_{t-1})},1\right\}. \label{e:ap}
	\end{equation}	
That is, we set
	\begin{equation}\label{e:assign}
		\mathbf{x}_{t}^j = \left\{
		\begin{aligned}
			&\mathbf{x}^{\ast}_{t}, ~\text{with probability} ~ a_t(\mathbf{x}^{\ast}_{t}|\mathbf{x}^{j}_{t-1}) \\
			&\mathbf{x}^{j}_{t-1}, ~\text{otherwise.} 
		\end{aligned}
		\right.
	\end{equation}		
	Once a forward Kernel $K_t(\mathbf{x}_t|\mathbf{x}_{t-1})$ is chosen, one can determine an optimal choice of $L_{t-1}$ by:
	\begin{equation}
		\begin{aligned}\label{L_opt}
			L_{t-1}^{opt}(\mathbf{x}_{t-1}|\mathbf{x}_{t}) &=
			\frac{q_{t-1}(\mathbf{x}_{t-1})K_{t}(\mathbf{x}_t|{\mathbf{x}_{t-1}})}{q_{t}(\mathbf{x}_{t})}\\
			&= \frac{q_{t-1}(\mathbf{x}_{t-1})K_{t}(\mathbf{x}_t|{\mathbf{x}_{t-1}})}{\int q_{t-1}(\mathbf{x}_{t-1})K_{t}(\mathbf{x}_t|{\mathbf{x}_{t-1}})d\mathbf{x}_{t-1}},
		\end{aligned}
	\end{equation}
where the optimality is achieved through yielding the minimal estimator variance \cite{del2006sequential}.
In reality, this optimal backward kernel usually cannot be used directly as the 
integral on the denominator cannot be calculated analytically. 
However, when the MCMC kernel is used, an approximate optimal kernel can be derived from Eq.~\eqref{L_opt}:
	\begin{equation}
		\label{L_mcmc}
			L_{t-1}(\mathbf{x}_{t-1}|\mathbf{x}_{t}) =
			\frac{q_{t}(\mathbf{x}_{t-1})K_{t}(\mathbf{x}_t|{\mathbf{x}_{t-1}})}{q_{t}(\mathbf{x}_{t})},
	\end{equation}
the detailed derivation can be found in \cite{del2006sequential}.
When Eq.~\eqref{L_mcmc} is used, the incremental weight function $\alpha_t(\mathbf{x}_{t-1},\mathbf{x}_t)$ in Eq. \ref{SMCS_weight_split}, reduces to the following:
	\begin{equation}
		\alpha_t(\mathbf{x}_{t-1},\mathbf{x}_t)=\frac{q_t(\mathbf{x}_{t-1})}{q_{t-1}(\mathbf{x}_{t-1})}.
	\label{eqAlphaMCMC}
	\end{equation}
Note that, interestingly only the previous sample is used in the weight calculation when Eq.~\eqref{L_mcmc} is used.
In our method, we use the MCMC kernel and Eq.~\eqref{L_mcmc} as the forward and backward kernels respectively. 

To alleviate sample degeneracy, a key step in SMCS is the resampling of samples according to their associated weights. 
The resampling algorithms are well documented, e.g. \cite{douc2005comparison}, and are not discussed here.	
In SMCS, typically resampling is conducted when the effective samples size (ESS) \cite{doucet2009tutorial} is lower than a prescribed threshold value $ESS_{\min}$. 
To conclude, we provide the complete procedure of SMCS in Algorithm \ref{alg:SMCS}, to generate $N$ samples from the target distribution $q_{t}(\cdot)$.

	\begin{algorithm}
	    \caption{Sequential Monte Carlo Sampler}
		\label{alg:SMCS}
		\textbf{input}: weighted ensemble $\{(x_{t-1}^j,w_{t-1}^j)\}_{j=1}^N$
		
		\textbf{for} {$j=1$ to $N$}
		
		\begin{algorithmic}
			\State (a) draw $\mathbf{\textbf{x}}^*_{t}$ from $k(\cdot|\mathbf{\textbf{x}}^j_{t-1})$
			\State (b) calculate the acceptance probability $a(\-\textbf{x}^*_t,\-\textbf{x}_{t-1}^j)$ using Eq.~\eqref{e:ap}
			\State (c) determine $\textbf{x}^j_t$ using Eq.~\eqref{e:assign} and $a(\-\textbf{x}^*_t,\-\textbf{x}_{t-1}^j)$
			\State(d) calculate $\alpha_t^j$ using Eq.~\eqref{eqAlphaMCMC}	
			\State (e) compute $w^{j}_{t} = w^j_{t-1}\alpha^j_t$
		\end{algorithmic}
			\textbf{end for}
			
		 normalize the weights calculated
		 
		calculate ESS
		
		    \textbf{if} {$ESS < ESS_{\min}$}
			\begin{algorithmic}
		 \State resample the ensemble and set $w^j_t= 1/N$ for $j=1,...,N$
		\end{algorithmic}
		    \textbf{end if}

	\end{algorithm}

As one can see from Algorithm~\ref{alg:SMCS}, the SMCS algorithm is easily parallelizable, which is the main advantage over MCMC for our purposes. 
In addition, since SMCS is designed for sampling from a sequence of target distributions, it can naturally take advantage of the similarity between two successive target distributions, like the warped distributions in two consecutive iterations of MMC, which will be further demonstrated in Section~\ref{Section:MMCSMCS}.

\section{Multicanonical Sequential Monte Carlo Sampler} \label{Section:MMCSMCS}
Our proposed algorithm, termed as the \emph{Multicanonical Sequential Monte Carlo Sampler} (MSMCS) uses SMCS to generate the samples in each 
MMC iteration.
As has been shown in Section~\ref{SMCSDetail}, SMCS can naturally be used to generate samples from a sequence of target distributions and is therefore well suited for MMC, where the biasing distributions within each MMC iteration can be considered as a sequence of distributions.
Though the implementation seems straightforward, there are still some issues that need to be addressed with in the proposed MSMCS method.

In the standard MMC method, using MCMC (denoted by MMC-MCMC), the samples generated are unweighted and as such the update procedure for $\Theta$'s - determined by the proportion of samples landing in each bin - is based on the samples being unweighted. However, as SMCS produces weighted samples, we need to adapt the MMC procedure to account for this, by altering the update procedure for the Theta distributions. Specifically, we change how the value of $\hat{H}_{t,i}$ - the estimator of $P_i$ - is determined. The update procedure, when using unweighted samples, is determined by Eq.~\eqref{e:params}. 
When SMCS is used, the update procedure needs to be modified, specifically Eq.~\eqref{e:Hti} becomes, 
\begin{equation}
	\hat{H}_{t,i} = \sum^{N}_{j = 1} I_{D_i}(\mathbf{x}^j)\;w(\mathbf{x}^{j}).
	\end{equation}

Another issue is that, for SMC to be effective, two successive distributions cannot be too far apart from each other;
otherwise, the samples are very likely to be rejected in the Metropolis step. Within the MMC method, there is no guarantee that the IS distributions obtained in two successive iterations are close to each other.
For example, in our numerical experiments, we have observed that, for high-dimensional problems, such an issue appears frequently in the first MMC step, due to the difference in the initial distribution $q_{0}(\textbf{x})$ and subsequent target distribution $q_{1}(\textbf{x})$.

To address this issue, we propose including a simulated tempering process in the method. 
Namely, we introduce a set of intermediate distributions in between $q_t$ and $q_{t+1}$, which we can apply SMCS too. 
Note that the difference in the IS distributions, can be attributed to differences in the $\Theta$-functions (i.e. $\Theta_{t}(\-x)$ and $\Theta_{t+1}(\-x)$), as per Eq.~\eqref{e:qt}. 
We choose a strictly increasing sequence of scalars $\{\alpha_k\}_{k=1}^K$ with $\alpha_0=0$ and  $\alpha_K=1$, such that the intermediate $\Theta$-functions are 
\begin{equation}
    \Theta_{k}(\textbf{x}) = \alpha_k\; \Theta_{t+1}(\textbf{x}) + (1-\alpha_k) \; \Theta_{t}(\textbf{x}).
\end{equation}
It follows that the sequence of intermediate distributions $\{q_k\}_{k=0}^K$ can be defined accordingly via Eq.~\eqref{e:qt},
and we apply SMCS to this sequence of distributions ultimately yielding samples from the target distribution $q_{t+1}(\-x)$. 
One can see that when $q_t$ and $q_{t+1}$ are close to each other, SMCS can efficiently generate samples from $q_{t+1}$ via the forward kernel and the samples from $q_{t}$, so this tempering process is not needed. However, for two consecutive IS distributions that are far apart, we found that whilst introducing more intermediate steps increases the computational time for generating samples according to the next target distribution $q_{t+1}(\textbf{x})$, overall the MMC converges faster, offsetting this increased cost.
Therefore, in our algorithm, tempering is only triggered when certain prescribed conditions are satisfied (e.g. $\|\Theta_t(\-x)-\Theta_{t+1}(\-x)\|$ exceeds a threshold value).

We have presented the proposed MSMCS method in a MMC framework: namely, we want to implement MMC for a given problem, where the samples are drawn 
from the target distribution $q_t$ using SMCS. Alternatively, we can also understand the method from a SMCS perspective: 
that is,  the SMCS method is used 
in a particular problem where the sequence of distributions are constructed via MMC. 

\section{Numerical Examples} \label{Section:NumEx}
In this section, we provide four numerical examples of increasing complexity to demonstrate the performance of the proposed MSMCS algorithm. By complexity, we are referring to the dimensionality of the problem and the rarity of the performance parameter values. Each numerical example also demonstrates a different aspect of the advantages our proposed method has over MMC-MCMC.
	
\subsection{Chi-Square Distribution}
In the first example, we consider the Chi-square distribution, a continuous distribution with $k$-degrees of freedom, describing the distribution of a sum of squared random variables. In this example, we demonstrate that MMC can be used to reconstruct the Chi-square distribution with very low error compared to the true analytical distribution, using both MCMC and SMCS.

If $x_1,...,x_k$ are independent zero-mean Gaussian random variables, with unit variance, then the sum of their squares,
	\begin{equation}
		y = \sum_{i=1}^{k} x^2_i,
	\end{equation}
is distributed according to the Chi-square distribution with $k$ degrees of freedom, where we often use the notation: $y \sim \chi^{2}(k)$.
In this example, we construct the Chi-square distribution for $k=20$ degrees of freedom, where the  analytical form of the PDF is available.

In both MMC-MCMC and MSMCS, we use $20$ iterations with $5\times10^{3}$ samples per iteration, to allow for a fair comparison. Within each MMC-MCMC iteration, a single long chain of $5\times10^{3}$ samples with no burn-in period is used, so all samples are utilised.

\begin{figure}
		\centering
		\includegraphics[width=1\textwidth]{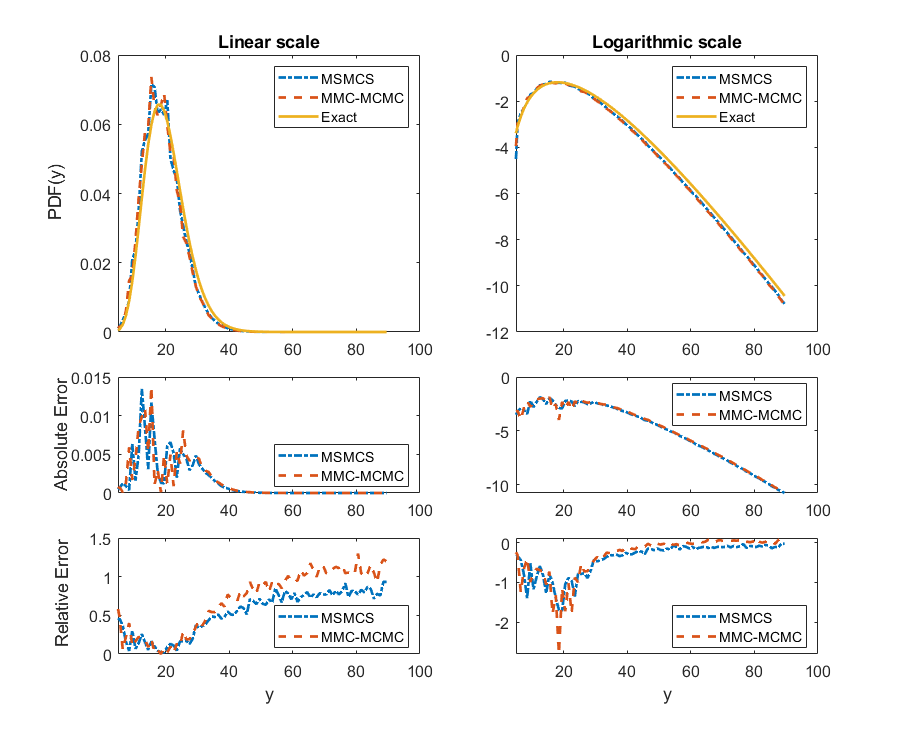}
		\caption{Chi-square distribution with 20 degrees of freedom computed by MSMCS and MMC-MCMC, compared to the analytical solution. The results are plotted on both the linear scale (left column) and the logarithmic scale (right column). The first row contains the approximated and analytical PDFs of y. The second and third rows show the absolute and relative errors of MMC compared to the analytical solution, respectively.}
		\label{fig:CSplot}
	\end{figure}

The results are shown in Figure \ref{fig:CSplot}, on both the linear and logarithmic scales. We also show the absolute and relative errors compared to the true analytical solution. 
The results demonstrate that the MMC method can reconstruct the Chi-square PDF with a low relative error compared to the true analytical solution, and that the MMC method can effectively explore the low probability events with a relatively small total sample size. In addition, the results show
that, in this relatively simple example, both the MSMCS and MMC-MCMC methods obtain comparable performance with regard to the error measures.

\subsection{Cantilever Beam Problem}
We now consider a real-world engineering example: a cantilever beam model studied in \cite{li2011efficient,wu1990advanced}. In this example, we impose a burn-in period on MCMC, as is often required, to ensure all the samples generated by MCMC follow the MMC distribution in each iteration. As outlined previously, this is not required for SMCS, where all samples can be utilised.

\begin{figure}
		\centering
		\includegraphics[width=0.55\textwidth]{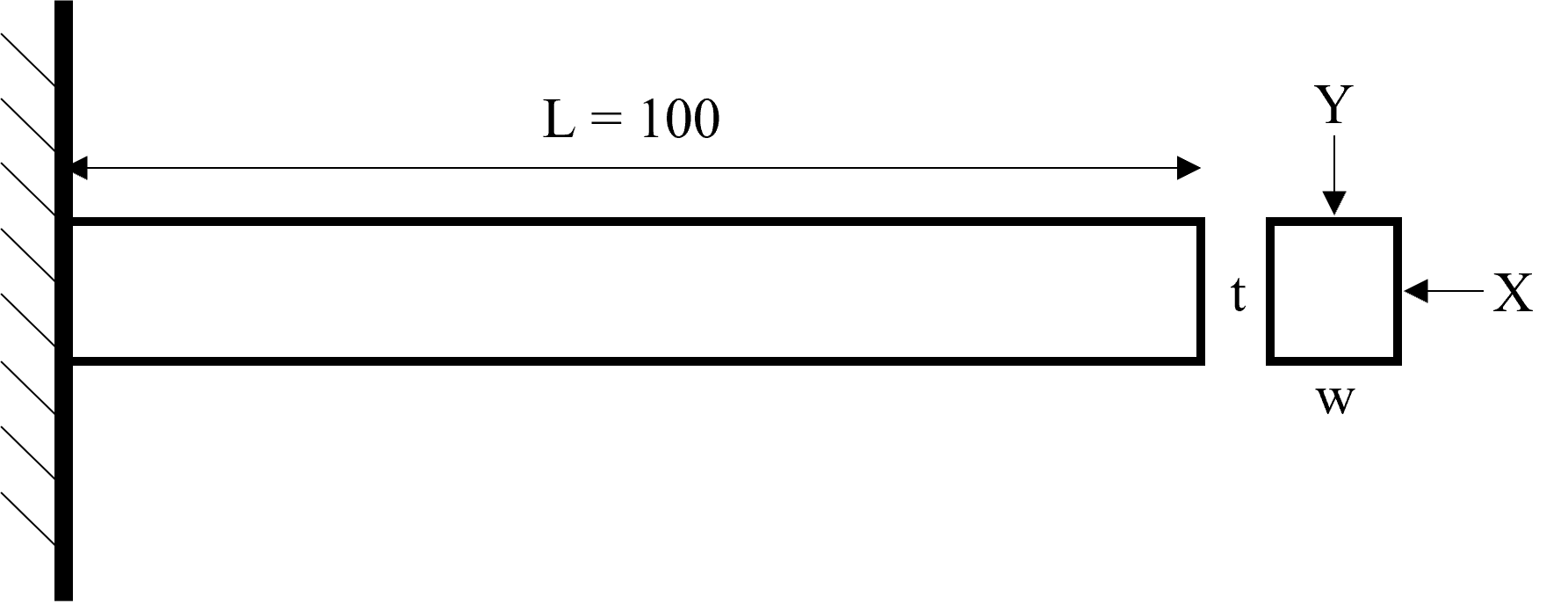}
		\caption{Cantilever Beam Problem}
		\label{fig:CantBeam}		
	\end{figure}

As illustrated in Figure \ref{fig:CantBeam}, we define our beam with width $w$, height $t$, length $L$, and elasticity $E$. We are interested in the beam's reliability when subjected to transverse load $Y$ and horizontal load $X$. This is a widely adopted testbed problem in reliability analysis, where the failure of the system relates to the maximum deflection of the beam $(y)$, as determined by the following equation:
	\begin{equation}
		y = \frac{4L^{3}}{Ewt} \sqrt{\left(\frac{Y}{t^{2}}\right)^{2}+\left(\frac{X}{w^{2}}\right)^{2}}
	\end{equation} 
	
Following the problem set up of \cite{li2011efficient,wu1990advanced}, we assume that the beam is of fixed length $L = 100$, with beam width $w$, height $t$, applied loads $X$ and $Y$, and elastic modulus of the material $E$ being random parameters, which are all independently distributed following a normal distribution. The mean and variance of each normally distributed parameter are provided in Table \ref{CantileverTable}.
	
	\begin{table}[ht]
		\caption{The mean and variance of the random parameters}
		\centering	
		\begin{tabular}{ cccccc } 
			\hline
			Parameter & $w$ & $t$ & $X$ & $Y$ & $E$ \\
			\hline
			Mean & $4$ & $4$ & $500$ & $1000$ & $2.9 \times 10^{6}$ \\ 
			Variance & $0.001$ & $0.0001$ & $100$ & $100$ & $1.45 \times 10^{6}$ \\ 
			\hline
		\end{tabular}
		\label{CantileverTable}
	\end{table}

We compute the PDF of $y$ with three methods: plain MC, MMC-MCMC and MSMCS. In the MC simulation, we use $10^{8}$ full model evaluations. In both MMC-MCMC and MSMCS, we use $20$ iterations with $5\times10^{4}$ samples in each iteration, to allow for a fair comparison. Within each MMC-MCMC iteration, we use a single long chain MCMC, and as such it cannot be implemented in parallel.  
Also in this example, we impose a burn-in period of $15\%$. We set $R_y = [5.35,6.80]$ divided into 145 bins, each of width 0.01.

\begin{figure}
		\centering
		\includegraphics[width=115mm]{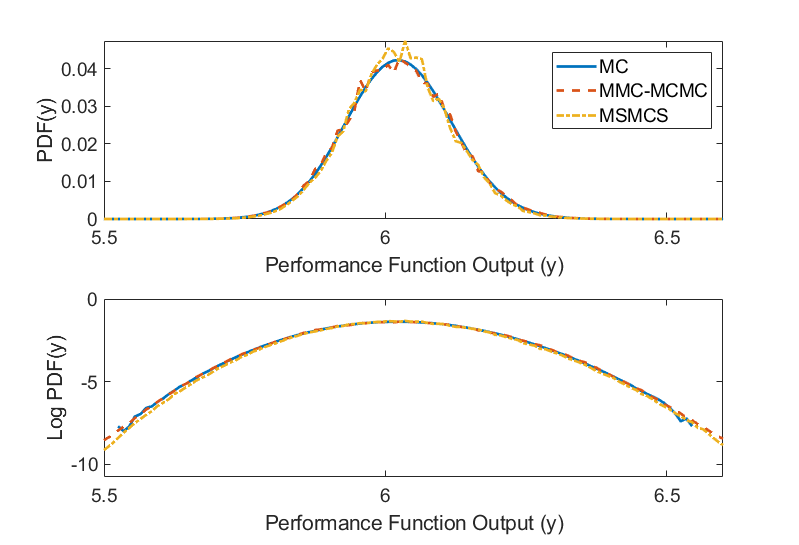}
		\caption{Cantilever Beam PDF computed by MC, MSMCS and MMC-MCMC. The results are shown on both the linear scale (left column) and logarithmic scale (right column).}
		\label{fig:CBplot}
	\end{figure}

To compare the results, we plot the PDF obtained by the three methods in Fig. \ref{fig:CBplot}.
First, one can see that the results of three methods agree very well in the high probability region, indicating that all the methods can correctly reproduce the sought PDF.
The two MMC based methods are substantially more effective in  the low probability regions -- the plain MC cannot reach
the same level of rarity (e.g. at $y=6.6$)  
while using 100 times more samples. 
The two MMC methods yield comparable results in this example, but as has been mentioned, MSMCS has the advantage of 
parallel implementation.
	
\subsection{Quarter Car Model}
In our third example, we consider a further real-world example: a quarter car model studied by Wong et al \cite{wong2008theory}. In this example, we implement MMC-MCMC in two alternate ways, to demonstrate the computational efficiency gained by using MSMCS - \emph{see implementation details}.

\subsubsection*{Problem Set Up}
\begin{figure}
		\centering
		\includegraphics[width=0.35\textwidth]{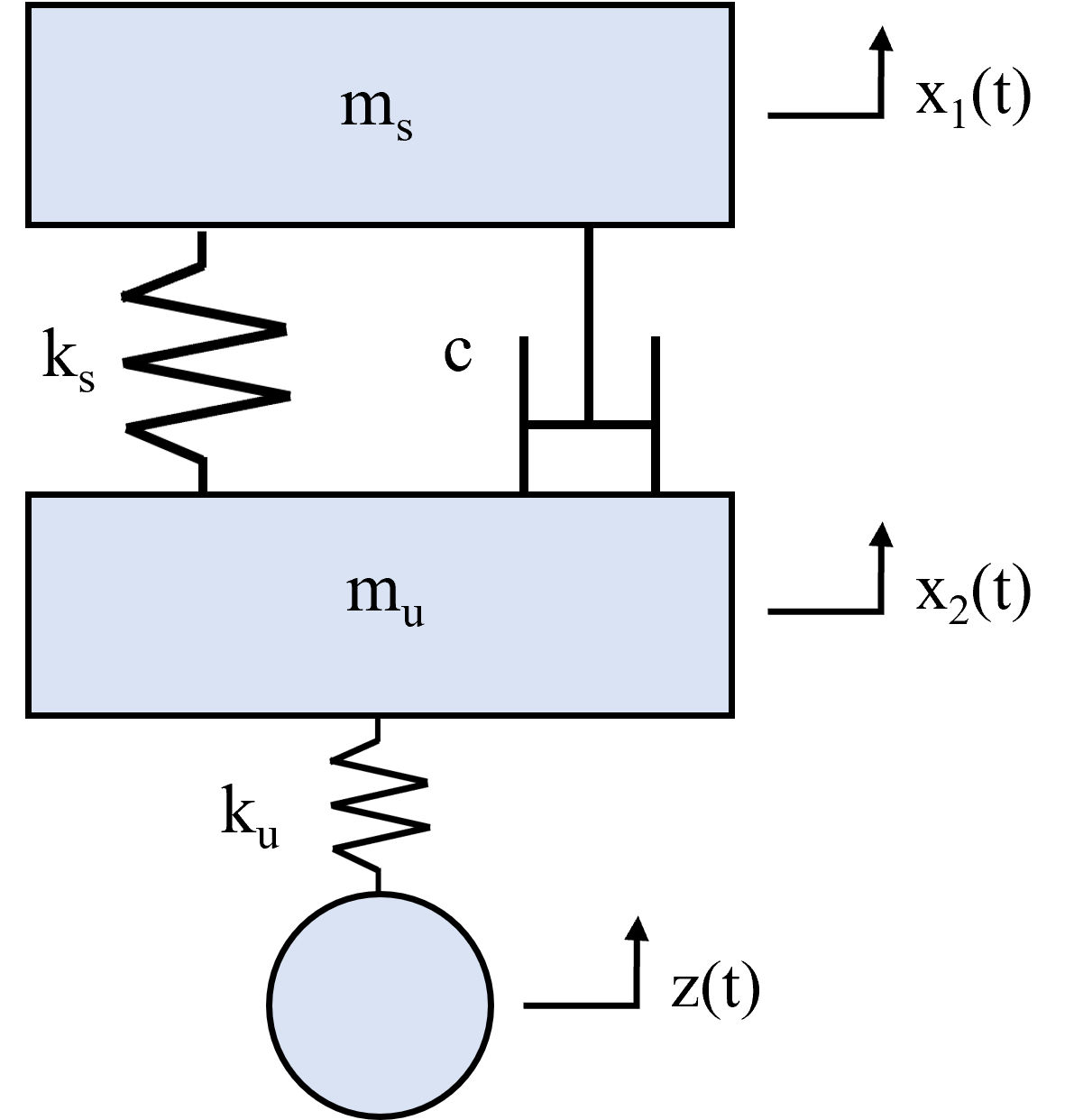}
		\caption{Quarter Car Model}
		\label{fig:QC}
	\end{figure}

The quarter-car model is used for vehicle suspension systems to investigate how they respond under a random road profile. As illustrated in Figure \ref{fig:QC}, we set-up our model following \cite{wong2008theory}, such that the sprung mass $m_{s}$ and the unsprung mass $m_{u}$ are connected by a non-linear spring (with stiffness $k_{s}$) and a linear damper (with damping coefficient $c$). The unsprung mass interacts with the road surface via a non-linear spring (with stiffness $k_{u}$). The displacement of the wheel $z(t)$ represents the interaction of the quarter car system with the road surface.

The displacements of the sprung and the unsprung masses are denoted by $x_{1}$ and $x_{2}$ respectively. Mathematically, the model is described by a two-degree-of-freedom ordinary differential equation (ODE) system:
	\begin{subequations}
		\begin{eqnarray}
			m_{s}\frac{d^{2}x_1}{dt^2} = -k_{s}(x_{1}-x_{2})^{3} - c\left(\frac{dx_1}{dt}-\frac{dx_2}{dt}\right),
		\end{eqnarray}
		\begin{equation}
			m_{u}\frac{d^{2}x_2}{dt^2} = k_{s}(x_{1}-x_{2})^{3} + c\left(\frac{dx_1}{dt}-\frac{dx_2}{dt}\right) +k_{u}(z(t)-x_{2}).
		\end{equation}
		\label{e:qcm}
	\end{subequations}
	
In our problem, the uncertainty arises through the random road profile $z(t)$ which is modelled as a zero-mean white Gaussian random force with standard deviation $\sigma = 1$. For the sake of our model, all other parameters are assumed to be fixed, taking the values as given by Table \ref{t:qcm}.
	
	\begin{table}[ht]
		\caption{The parameter values of the quarter car model}
		\centering	
		\begin{tabular}{ccccc}
			\hline
			$m_{s}$ & $m_{u}$ & $k_{s}$ & $k_{u}$ & $c$ \\
			\hline
			$20$ & $40$ & $400$ & $2000$ & $600$ \\ 
			\hline
		\end{tabular}
		\label{t:qcm}
	\end{table}
	
We are interested in the maximum difference between the displacements of the sprung and unsprung springs in a given interval $[0, T]$, as calculated by:
	\begin{equation}
		y = \underset{0 \leq t \leq T}{\mbox{max}} \{|x_{1}(t) - x_{2}(t)|\}.
	\end{equation}	
In extreme scenarios when this displacement exceeds a certain value, say $y^{\ast}$, the car's suspension would break. We want to reconstruct the entire probability density function (PDF) of y. With the PDF, we can estimate the probability $\PR(y > y^{\ast})$ for any value of $y^{\ast}$ in the range of interest.

\subsubsection*{Implementation Details}
We solve Eqs. \ref{e:qcm} numerically using the 4-th order Runge-Kutta method where the step size is taken to be $\Delta t = T/100$, so the random variable in this problem is effectively of 100 dimensions. We take $T=1$ and set initial conditions of Eqs. \ref{e:qcm} to be 
	\begin{equation}
		x_{1}(0) = \frac{dx_1}{dt}(0) = 0,\;\; x_{2}(0) = \frac{dx_2}{dt}(0) = 0
	\end{equation}
	
We conduct a standard MC simulation with $10^{6}$ samples. In both MSMCS and MMC-MCMC, we use $20$ iterations with $2\times10^{4}$ samples in each iteration. The MSMCS method is easily parallelisable, meaning that within each MMC iteration, one can update the new samples completely in parallel according to the target MMC distribution, rather than forming a single long chain - significantly improving the computational efficiency. To provide a fair computational comparison, for this example, we conduct MMC-MCMC in two ways. In the first case, we use a single long chain of length $2\times10^{4}$ - the most typical implementation of MCMC, which is also how the MCMC is implemented in the first two examples. In the second case, within each iteration we use $10$ chains each of length $2\times10^{3}$, to provide a fairer comparison to the parallel implementation of MSMCS.

\subsubsection*{Results}
\begin{figure}
	\centering
	\includegraphics[width=.85\textwidth]{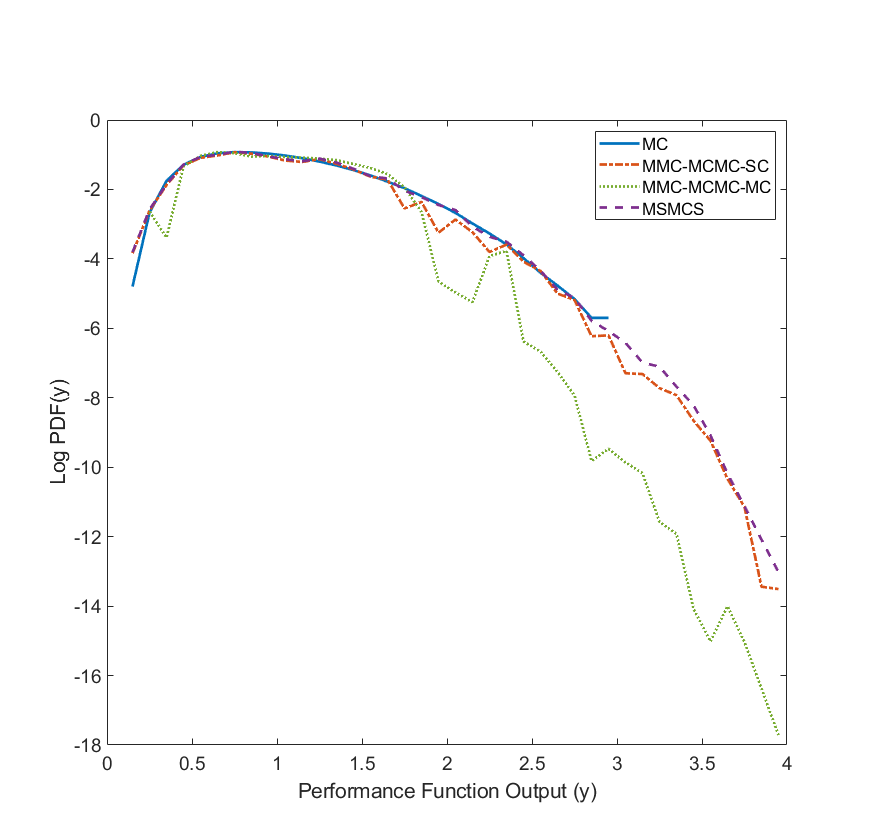}
	\caption{Quarter Car Model PDF computed by MC, MSMCS and MMC-MCMC. MMC-MCMC-SC uses a single long chain. MMC-MCMC-MC uses ten shorter chains in parallel. The results are shown on both the linear scale (left column) and the logarithmic scale (right column).}
	\label{fig:QCplot}
\end{figure}

The results of all three methods are shown in Figure \ref{fig:QCplot}. The MC method only estimated the PDF to the order of $10^{-6}$ (as expected), while the MSMCS method estimated it to order $10^{-12}$. MMC-MCMC with a single chain (referred to as MMC-MCMC-SC), also accurately reconstructed the performance parameter PDF, however MMC-MCMC with multiple chains (referred to as MMC-MCMC-MC) and therefore enabling parallel implementation, significantly underestimated the PDF values for values $y>1.8$.
The results indicate that due to the sequential nature of MCMC, running multiple short chains  substantially undermines the performance of the method. 
Therefore, on the basis of parallel implementation,  the MSMCS method clearly outperforms MMC-MCMC.

\subsection{Copula Model}
The development of rare event simulation techniques is also critical for 
the risk management in financial markets. Therefore, the final application we investigate is applying the MMC method to a Copula model - one of the most widely used portfolio risk models. A copula model allows one to separate the dependence structure of the portfolio from the marginal densities of each variables - representing the individual risks of each obligor - which can have different probability distributions. We consider the Student's t-copula model, proposed by Bassamboo 
\textit{et al} \cite{bassamboo2008portfolio}.

\subsubsection*{Problem Set Up}
We follow the problem set up of \cite{bassamboo2008portfolio} and \cite{chan2010efficient}. Consider a portfolio of loans consisting of $n$ obligors, we aim to find the distribution of losses from defaults over a fixed time horizon, from which we can determine large loss probabilities.
Suppose the probability of default for the $i$th obligor over the time horizon is $p_i \in (0,1)$, for $i=1,...,n,$ and that in the event that the $i$th obligor defaults, a fixed and given loss of $c_i$ monetary units occurs. We begin by introducing a vector of underlying latent variables $\textbf{X} = (X_{1},...,X_{n})$ such that the $i$th obligor defaults if $X_i$ exceeds a given threshold level $x_i$. This threshold $x_i$ is set according to the marginal default probability of the $i$th asset, so that $\PR(X_i > x_i) = p_i$.

The portfolio loss from defaults is given by
\begin{equation}
    L(\textbf{X}) = c_{1}I_{\{X_1>x_1\}} + ... + c_{n}I_{\{X_n>x_n\}}
\end{equation}
where $I_{\{X_i>x_i\}}$ denotes the indicator function, which is equal to $1$ if $X_i > x_i$ and 0 otherwise.
We let the common risk factor and the individual idiosyncratic risks be independent normally distributed random variables, that is,
\begin{equation}
    Z \sim N(0,1)\;\text{and}\; \eta_i \sim N(0,\sigma^{2}_{\eta})\text{, for }i=1,...,n.
\end{equation}
We choose $0<p<1$ and let
\begin{equation}
    X_i = \frac{pZ + \sqrt{1-p^2}\eta_i}{T}, i=1,...,n,
\end{equation}
where $T$ is a non-negative random variable, independent of the other risk factors.

For a positive integer $k$, let $T = \sqrt{k^{-1}{\Gamma}(1/2,k/2)}$ where $\Gamma$ represents the PDF of the Gamma distribution \cite{bassamboo2008portfolio}. Therefore, our latent variables follow a multivariate t-distribution, whose dependence structure is given by a t-copula with $k$ degrees of freedom.

\subsubsection*{Implementation Details}
We use the same set up as Chan \textit{et al} \cite{chan2010efficient}, that is, we set $\sigma^{2}_{\eta} = 9$, $x = \sqrt{n}\; \text{x}\; 0.5$, $p = 0.25$, and $c=1$.
We conduct a standard MC simulation, with different sample sizes - as detailed in the results tables. In both MMC-MCMC and MSMCS, we use $20$ iterations with $1\times10^4$ samples in each iteration. We implement MMC-MCMC in twos forms, one with a single long chain - as it would typically be implemented - and one with parallel chains (100 chains each of length 100), which provides a fairer comparison to parallel implementation of MSMCS. Neither MCMC case uses a burn-in period.

\subsubsection*{Results}
We are interested in the probability of large losses, defined as the loss function value $L(\textbf{X}) > l$, where $l = bn$ for different samples sizes $n$ and different threshold values $b$. We vary either the degrees of freedom $k$ or the sample size $n$, and for each of these scenarios, we determine the probability that the loss exceeds $l = b \textbf{x} n$, for $b=0.1,0.2,0.25,0.3$. The results are presented in Table \ref{Table:CopRes}.

\begin{table}
\caption{Copula Results using MC; MSMCS; and MMC-MCMC.}
\label{Table:CopRes}

\begin{subtable}[t]{\textwidth}
\caption{$k=4$  \&  $n=250$}
\resizebox{\columnwidth}{!}{%
\begin{tabular}{|l|l|llll|}
\hline
\multicolumn{1}{|c|}{\multirow{2}{*}{\begin{tabular}[c]{@{}c@{}}Large Loss \\ Threshold (b)\end{tabular}}} &
  \multicolumn{1}{c|}{Sample Size} &
  \multicolumn{4}{c|}{Probability Estimate} \\ \cline{2-6} 
\multicolumn{1}{|c|}{} &
  \multicolumn{1}{c|}{MC} &
  \multicolumn{1}{c|}{MC} &
  \multicolumn{1}{c|}{MMC-MCMC-SC} &
  \multicolumn{1}{c|}{MMC-MCMC-MC} &
  \multicolumn{1}{c|}{MSMCS} \\ \hline
$0.1$  & $5 \times 10^{5}$ & \multicolumn{1}{l|}{$7.36 \times 10^{-2}$} & \multicolumn{1}{l|}{$7.27 \times 10^{-2}$} & \multicolumn{1}{l|}{$1.69 \times 10^{-1}$} & \multicolumn{1}{l|}{$7.31 \times 10^{-2}$} \\ \hline
0.2  & $5 \times 10^{5}$ & \multicolumn{1}{l|}{$1.72 \times 10^{-2}$} & \multicolumn{1}{l|}{$1.63 \times 10^{-2}$} & \multicolumn{1}{l|}{$5.96 \times 10^{-2}$}  & $1.71 \times 10^{-2}$ \\ \hline
0.25 & $5 \times 10^{5}$ & \multicolumn{1}{l|}{$8.08 \times 10^{-3}$} & \multicolumn{1}{l|}{$8.13 \times 10^{-3}$} & \multicolumn{1}{l|}{$3.29 \times 10^{-2}$}  &  $8.05 \times 10^{-3}$ \\ \hline
0.3  & $5 \times 10^{5}$ & \multicolumn{1}{l|}{$3.21 \times 10^{-3}$} & \multicolumn{1}{l|}{$3.24 \times 10^{-3}$} & \multicolumn{1}{l|}{$1.71 \times 10^{-2}$}  & $3.28 \times 10^{-3}$ \\ \hline
\end{tabular}%
}
\vspace{0.1cm}
\end{subtable}

\begin{subtable}[t]{\textwidth}
\caption{$k=8$ \&  $n=250$}
\resizebox{\columnwidth}{!}{%
\begin{tabular}{|l|l|llll|}
\hline
\multicolumn{1}{|c|}{\multirow{2}{*}{\begin{tabular}[c]{@{}c@{}}Large Loss \\ Threshold (b)\end{tabular}}} &
  \multicolumn{1}{c|}{Sample Size} &
  \multicolumn{4}{c|}{Probability Estimate} \\ \cline{2-6} 
\multicolumn{1}{|c|}{} &
  \multicolumn{1}{c|}{MC} &
  \multicolumn{1}{c|}{MC} &
  \multicolumn{1}{c|}{MMC-MCMC-SC} &
  \multicolumn{1}{c|}{MMC-MCMC-MC} &
  \multicolumn{1}{c|}{MSMCS} \\ \hline
0.1  & $5 \times 10^{6}$ & \multicolumn{1}{l|}{$1.45 \times 10^{-2}$} & \multicolumn{1}{l|}{$1.39 \times 10^{-2}$} & \multicolumn{1}{l|}{$2.24 \times 10^{-3}$}  & \multicolumn{1}{l|}{$1.42 \times 10^{-2}$} \\ \hline
0.2  & $5 \times 10^{6}$ & \multicolumn{1}{l|}{$9.49 \times 10^{-4}$} & \multicolumn{1}{l|}{$9.43 \times 10^{-4}$} & \multicolumn{1}{l|}{$1.66 \times 10^{-4}$}  & $9.49 \times 10^{-4}$ \\ \hline
0.25 & $5 \times 10^{6}$ & \multicolumn{1}{l|}{$2.38 \times 10^{-4}$} & \multicolumn{1}{l|}{$2.49 \times 10^{-4}$} & \multicolumn{1}{l|}{$4.29 \times 10^{-5}$}  & $2.46 \times 10^{-4}$ \\ \hline
0.3  & $5 \times 10^{6}$ & \multicolumn{1}{l|}{$4.04 \times 10^{-5}$} & \multicolumn{1}{l|}{$3.98 \times 10^{-5}$} & \multicolumn{1}{l|}{$1.04 \times 10^{-5}$}  & $4.01 \times 10^{-5}$ \\ \hline
\end{tabular}%
}
\vspace{0.1cm}
\end{subtable}

\begin{subtable}[t]{1\textwidth}
\caption{$k=12$ \& $n=250$}
\resizebox{\columnwidth}{!}{%
\begin{tabular}{|l|l|llll|}
\hline
\multicolumn{1}{|c|}{\multirow{2}{*}{\begin{tabular}[c]{@{}c@{}}Large Loss \\ Threshold (b)\end{tabular}}} &
  \multicolumn{1}{c|}{Sample Size} &
  \multicolumn{4}{c|}{Probability Estimate} \\ \cline{2-6} 
\multicolumn{1}{|c|}{} &
  \multicolumn{1}{c|}{MC} &
  \multicolumn{1}{c|}{MC} &
  \multicolumn{1}{c|}{MMC-MCMC-SC} &
  \multicolumn{1}{c|}{MMC-MCMC-MC} &
  \multicolumn{1}{c|}{MSMCS} \\ \hline
0.1  & $5 \times 10^{7}$ & \multicolumn{1}{l|}{$9.77 \times 10^{-3}$} & \multicolumn{1}{l|}{$9.82 \times 10^{-3}$} & \multicolumn{1}{l|}{$5.96 \times 10^{-5}$}  & $9.78 \times 10^{-3}$ \\ \hline
0.2  & $5 \times 10^{7}$ & \multicolumn{1}{l|}{$7.49 \times 10^{-3}$} & \multicolumn{1}{l|}{$7.63 \times 10^{-3}$} & \multicolumn{1}{l|}{$1.04 \times 10^{-6}$}  & $7.53 \times 10^{-3}$ \\ \hline
0.25 & $5 \times 10^{7}$ & \multicolumn{1}{l|}{$1.05 \times 10^{-5}$} & \multicolumn{1}{l|}{$1.02 \times 10^{-5}$} & \multicolumn{1}{l|}{$1.22 \times 10^{-7}$}  & $1.03 \times 10^{-5}$ \\ \hline
0.3  & $5 \times 10^{7}$ & \multicolumn{1}{l|}{$1.12 \times 10^{-6}$} & \multicolumn{1}{l|}{$1.34 \times 10^{-6}$} & \multicolumn{1}{l|}{$1.65 \times 10^{-8}$}  & $1.21 \times 10^{-6}$ \\ \hline
\end{tabular}%
}
\vspace{0.1cm}
\end{subtable}

\begin{subtable}[t]{1\textwidth}
\caption{$k=16$ \& $n=250$}
\resizebox{\columnwidth}{!}{%
\begin{tabular}{|l|l|llll|}
\hline
\multicolumn{1}{|c|}{\multirow{2}{*}{\begin{tabular}[c]{@{}c@{}}Large Loss \\ Threshold (b)\end{tabular}}} &
  \multicolumn{1}{c|}{Sample Size} &
  \multicolumn{4}{c|}{Probability Estimate} \\ \cline{2-6} 
\multicolumn{1}{|c|}{} &
  \multicolumn{1}{c|}{MC} &
  \multicolumn{1}{c|}{MC} &
  \multicolumn{1}{c|}{MMC-MCMC-SC} &
  \multicolumn{1}{c|}{MMC-MCMC-MC} &
  \multicolumn{1}{c|}{MSMCS} \\ \hline
0.1  & $5 \times 10^{8}$ & \multicolumn{1}{l|}{$9.40 \times 10^{-4}$} & \multicolumn{1}{l|}{$9.36 \times 10^{-4}$} & \multicolumn{1}{l|}{$2.50 \times 10^{-6}$}  & $9.43 \times 10^{-4}$ \\ \hline
0.2  & $5 \times 10^{8}$ & \multicolumn{1}{l|}{$6.91 \times 10^{-6}$} & \multicolumn{1}{l|}{$6.90 \times 10^{-6}$} & \multicolumn{1}{l|}{$9.58 \times 10^{-9}$}  & $6.86 \times 10^{-6}$ \\ \hline
0.25 & $5 \times 10^{8}$ & \multicolumn{1}{l|}{$6.22 \times 10^{-7}$} & \multicolumn{1}{l|}{$6.18 \times 10^{-7}$} & \multicolumn{1}{l|}{$6.04 \times 10^{-10}$}  & $6.19 \times 10^{-7}$ \\ \hline
0.3  & $5 \times 10^{8}$ & \multicolumn{1}{l|}{$4.40 \times 10^{-8}$} & \multicolumn{1}{l|}{$4.37 \times 10^{-8}$} & \multicolumn{1}{l|}{$3.67 \times 10^{-11}$}  & $4.51 \times 10^{-8}$ \\ \hline
\end{tabular}%
}
\vspace{0.1cm}
\end{subtable}

\begin{subtable}[t]{1\textwidth}
\caption{$k=20$ \& $n=250$}
\resizebox{\columnwidth}{!}{%
\begin{tabular}{|l|l|llll|}
\hline
\multicolumn{1}{|c|}{\multirow{2}{*}{\begin{tabular}[c]{@{}c@{}}Large Loss \\ Threshold (b)\end{tabular}}} &
  \multicolumn{1}{c|}{Sample Size} &
  \multicolumn{4}{c|}{Probability Estimate} \\ \cline{2-6} 
\multicolumn{1}{|c|}{} &
  \multicolumn{1}{c|}{MC} &
  \multicolumn{1}{c|}{MC} &
  \multicolumn{1}{c|}{MMC-MCMC-SC} &
  \multicolumn{1}{c|}{MMC-MCMC-MC} &
  \multicolumn{1}{c|}{MSMCS} \\ \hline
0.1  & $5 \times 10^{8}$ & \multicolumn{1}{l|}{$2.83 \times 10^{-4}$} & \multicolumn{1}{l|}{$2.88 \times 10^{-4}$} & \multicolumn{1}{l|}{$1.39 \times 10^{-7}$}  & $2.76 \times 10^{-4}$ \\ \hline
0.2  & $5 \times 10^{8}$ & \multicolumn{1}{l|}{$7.98 \times 10^{-7}$} & \multicolumn{1}{l|}{$7.61 \times 10^{-7}$} & \multicolumn{1}{l|}{$1.35 \times 10^{-10}$}  & $7.73 \times 10^{-7}$ \\ \hline
0.25 & $5 \times 10^{8}$ & \multicolumn{1}{l|}{$5.40 \times 10^{-8}$} & \multicolumn{1}{l|}{$4.92 \times 10^{-8}$} & \multicolumn{1}{l|}{$2.99 \times 10^{-12}$}  & $5.32 \times 10^{-8}$ \\ \hline
0.3  & $5 \times 10^{8}$ & \multicolumn{1}{l|}{0} & \multicolumn{1}{l|}{$5.72 \times 10^{-9}$} & \multicolumn{1}{l|}{$1.02 \times 10^{-13}$}  & $5.63 \times 10^{-9}$ \\ \hline
\end{tabular}%
}
\vspace{0.1cm}
\end{subtable}

\begin{subtable}[t]{1\textwidth}
\caption{$k=12$ \& $n=500$}
\resizebox{\columnwidth}{!}{%
\begin{tabular}{|l|l|llll|}
\hline
\multicolumn{1}{|c|}{\multirow{2}{*}{\begin{tabular}[c]{@{}c@{}}Large Loss \\ Threshold (b)\end{tabular}}} &
  \multicolumn{1}{c|}{Sample Size} &
  \multicolumn{4}{c|}{Probability Estimate} \\ \cline{2-6} 
\multicolumn{1}{|c|}{} &
  \multicolumn{1}{c|}{MC} &
  \multicolumn{1}{c|}{MC} &
  \multicolumn{1}{c|}{MMC-MCMC-SC} &
  \multicolumn{1}{c|}{MMC-MCMC-MC} &
  \multicolumn{1}{c|}{MSMCS} \\ \hline
0.1  & $5 \times 10^{8}$ & \multicolumn{1}{l|}{$9.61 \times 10^{-5}$} & \multicolumn{1}{l|}{$9.42 \times 10^{-5}$} & \multicolumn{1}{l|}{$5.08 \times 10^{-12}$} & $9.52 \times 10^{-5}$ \\ \hline
0.2  & $5 \times 10^{8} $& \multicolumn{1}{l|}{$1.34 \times 10^{-6}$} & \multicolumn{1}{l|}{$1.39 \times 10^{-6}$} & \multicolumn{1}{l|}{$7.15 \times 10^{-13}$} & $1.38 \times 10^{-6}$ \\ \hline
0.25 & $5 \times 10^{8}$ & \multicolumn{1}{l|}{$1.36 \times 10^{-7}$} & \multicolumn{1}{l|}{$1.57 \times 10^{-7}$} & \multicolumn{1}{l|}{$4.37 \times 10^{-13}$} & $0.84 \times 10^{-7}$ \\ \hline
0.3  & $5 \times 10^{8}$ & \multicolumn{1}{l|}{$1.00 \times 10^{-8}$} & \multicolumn{1}{l|}{$1.29 \times 10^{-8}$} & \multicolumn{1}{l|}{$2.54 \times 10^{-13}$} & $1.27 \times 10^{-8}$ \\ \hline
\end{tabular}%
}
\vspace{0.1cm}
\end{subtable}

\begin{subtable}[t]{1\textwidth}
\caption{$k=12$ \&  $n=1000$}
\resizebox{\columnwidth}{!}{%
\begin{tabular}{|l|l|llll|}
\hline
\multicolumn{1}{|c|}{\multirow{2}{*}{\begin{tabular}[c]{@{}c@{}}Large Loss \\ Threshold (b)\end{tabular}}} &
  \multicolumn{1}{c|}{Sample Size} &
  \multicolumn{4}{c|}{Probability Estimate} \\ \cline{2-6} 
\multicolumn{1}{|c|}{} &
  \multicolumn{1}{c|}{MC} &
  \multicolumn{1}{c|}{MC} &
  \multicolumn{1}{c|}{MMC-MCMC-SC} &
  \multicolumn{1}{c|}{MMC-MCMC-MC} &
  \multicolumn{1}{c|}{MSMCS} \\ \hline
0.1  & $3 \times 10^{8}$ & \multicolumn{1}{l|}{$1.96 \times 10^{-6}$} & \multicolumn{1}{l|}{$1.88 \times 10^{-6}$} & \multicolumn{1}{l|}{$2.54 \times 10^{-13}$} & $1.91 \times 10^{-6}$  \\ \hline
0.2  & $3 \times 10^{8}$ & \multicolumn{1}{l|}{$3.67 \times 10^{-8}$} & \multicolumn{1}{l|}{$3.58 \times 10^{-8}$} & \multicolumn{1}{l|}{$6.29 \times 10^{-14}$} & $3.72 \times 10^{-8}$  \\ \hline
0.25 & $3 \times 10^{8}$ & \multicolumn{1}{l|}{$2.39 \times 10^{-9}$} & \multicolumn{1}{l|}{$2.24 \times 10^{-9}$}  & \multicolumn{1}{l|}{$4.18 \times 10^{-14}$} & $2.28 \times 10^{-9}$  \\ \hline
0.3  & $3 \times 10^{8}$ & \multicolumn{1}{l|}{0} & \multicolumn{1}{l|}{$3.25 \times 10^{-10}$}  & \multicolumn{1}{l|}{$7.24 \times 10^{-15}$} & $3.19 \times 10^{-10}$ \\ \hline
\end{tabular}%
}

\end{subtable}
\end{table}

As the MMC method reconstructs the whole loss distribution, we only require seven simulations to be performed, from which the loss probability for any $b$-value can be obtained. This is a significant computational saving, compared to with other existing methods, like the Conditional-MC in \cite{chan2010efficient}, which would require a new simulation for each $b$-value. Our results show that the MMC method - with both MCMC and SMCS - produces significant computational savings for estimating large loss probabilities, given a Copula model. Both MMC-MCMC (with a single long chain, denoted by MMC-MCMC-SC) and MSMCS, are very effective here - although, MMC-MCMC (with multiple parallel chains, denoted by MMC-MCMC-MC) performs poorly, particularly in the high-dimensional setting,
clear illustrating  the advantage of MSMCS in the parallel implementation. 
 Finally, as shown by comparison to the standard MC, MMC is very effective method for the purposes of a Copula model and estimating large loss probabilities.

\section{Conclusion} \label{Section:Conclusion}
In summary, we consider UQ problems where the full distribution of a performance parameter is sought, and we propose a method to do so by incorporating the MMC and SMCS methods.
Specifically the method uses SMCS instead of MCMC to draw samples from the warped distributions in each iteration of MMC. We have demonstrated that the proposed MSMCS method can outperform both the standard MMC-MCMC, in the sense that SMCS is easily parallelisable and so it can take full advantage of parallel high-powered computing, while MCMC, due to its sequential nature, requires a (often very long) burn-in period, which in fact is the reason that the implementation with multiple short chains does not perform well. We believe that our proposed algorithm has wide applicability, improving the computational efficiency associated with finding failure probabilities or reconstructing the whole probability distribution of interest.

One weakness of the proposed method is that MCMC is easier to implement than SMCS and involves simpler computations - so MMC-MCMC is marginally faster than MSMCS to run. However, if one can use a parallel implementation then MSMCS significantly outperforms MMC-MCMC, as shown in the numerical examples. More importantly, both approaches to MMC can struggle in high-dimensional settings, where the generation of a new sample is likely to get rejected, which should be dealt with by developing and utilising more effective proposal distributions, for example, that based on the Hamiltonian dynamics~\cite{neal2011mcmc}.

\bibliography{MMC_SMCS}

\end{document}